\documentclass[aps,prb,superscriptaddress,showpacs,twocolumn]{revtex4}
\usepackage{graphicx}
\usepackage{dcolumn}
\usepackage{bm}

\begin{document}

\title{Quantum Jump Approach to Switching Process of a Josephson Junction
Coupled to a Microscopic Two-Level System}
\author{ Xueda Wen}
\affiliation{National Laboratory of Solid State Microstructures and Department of
Physics, Nanjing University, Nanjing 210093, China }
\author{ Yiwen Wang}
\affiliation{National Laboratory of Solid State Microstructures and Department of
Physics, Nanjing University, Nanjing 210093, China }
\author{ Ning Dong}
\affiliation{National Laboratory of Solid State Microstructures and Department of
Physics, Nanjing University, Nanjing 210093, China }
\author{ Guozhu Sun}
\affiliation{Research Institute of Superconductor Electronics and Department of
Electronic Science and Engineering, Nanjing University, Nanjing 210093,
People's Republic of China}
\author{ Jian Chen}
\affiliation{Research Institute of Superconductor Electronics and Department of
Electronic Science and Engineering, Nanjing University, Nanjing 210093,
People's Republic of China}
\author{ Lin Kang}
\affiliation{Research Institute of Superconductor Electronics and Department of
Electronic Science and Engineering, Nanjing University, Nanjing 210093,
People's Republic of China}
\author{ Weiwei Xu}
\affiliation{Research Institute of Superconductor Electronics and Department of
Electronic Science and Engineering, Nanjing University, Nanjing 210093,
People's Republic of China}
\author{ Peiheng Wu}
\affiliation{Research Institute of Superconductor Electronics and Department of
Electronic Science and Engineering, Nanjing University, Nanjing 210093,
People's Republic of China}
\author{Yang Yu}
\email{ yuyang@nju.edu.cn}
\affiliation{National Laboratory of Solid State Microstructures and Department of
Physics, Nanjing University, Nanjing 210093, China }

\begin{abstract}
With microwave irradiation, the switching current of a Josephson junction
coupled to a microscopic two-level system jumps randomly between two
discrete states. We modeled the switching process of the coupled system with
quantum jump approach that was generally used in quantum optics. The
parameters that affect the character of the quantum jumps between
macroscopic quantum states are discussed. The results obtained from our
theoretical analysis agree well with those of the experiments and provide a
clear physical picture for the macroscopic quantum jumps in Josephson
junctions coupled with two-level systems. In addition, quantum jumps may
serve as a useful tool to investigate the microscopic two-level structures
in solid-state systems.
\end{abstract}

\maketitle

\section{INTRODUCTION}

Recent progress on superconducting qubits based upon Josephson Junction (JJ)
unambiguously demonstrated the quantum behavior of the macroscopic variables.
\cite{Makhlin,Yu,Martinis,Yamamoto,Chiorescu,Oliver,Schuster}
Moreover, quantum jumps, an interesting quantum
phenomenon previously studied in quantum optics,\cite{Scully,Orszag,Plenio}
was experimentally demonstrated for the first time in a junction coupled
with a microscopic two-level system (TLS) recently.\cite{Yu2} The JJ-TLS
coupling system possesses $\Lambda $-type energy level structure and
microwave photons are used to generate transitions between quantum states.
However, the state of the system is read out by detecting macroscopic
quantum tunneling process rather than that by detecting photon emissions in
quantum optics. Quantum jumps then manifests itself in the form of jumping
randomly between upper branch and lower branch of the switching currents. In
the language of quantum measurement theory, the switching currents in the
upper branch or lower branch serve as a pointer from which the macroscopic
quantum state of the JJ-TLS coupling system can be determined. In this
situation the ensemble description of the dynamics of junctions based on the
master equation method \cite{Chow,Kopietz,Silvestrini} fails in describing
trajectories of a single quantum system. Since quantum jump approach
developed in the 1980s has made great successes in describing fluorescence
of single trapped ions\cite{Plenio}, in this paper we generalize the quantum
jump approach to the switching process of JJ-TLS coupling system and make a
systematic study of the parameters that have effects on the process. The
same method has also been used to investigate quantum jumps in Rabi
oscillations of a JJ-TLS coupling system.\cite{Wen} However, in that work
the biased current of the junction is fixed at an appropriate value while
here it keeps changing during the switching current measurement. Therefore,
new mechanisms such as Landau-Zener transitions may involve in the dynamics
of the JJ-TLS coupling system.

This article is organized as follows. In Sec. II, we describe the physics of
the current-biased Josephson junction briefly and introduce the quantum jump
approach for simulating the switching process of a current-biased junction.
In Sec. III we generalize the quantum jump approach to the switching process
of JJ-TLS coupling system and discuss the parameters that have effects on
the process. In Sec.IV we compare our theoretical results with experimental
data and make a conclusion in Sec.V.

\section{QUANTUM JUMP APPROACH TO SWITCHING PROCESS OF A CURRENT-BIASED
JOSEPHSON JUNCTION}

The Hamiltonian of a current-biased Josephson junction as shown in Fig.1(a)
reads\cite{Martinis2,Martinis3}
\begin{equation}  \label{Qubit Hamiltonion}
H_{JJ}=\frac{1}{2C}\hat{Q}^{2}-\frac{I_{0}\Phi_{0}}{2\pi}\cos\hat{\delta}-
\frac{I\Phi_{0}}{2\pi}\hat{\delta},
\end{equation}
where $I_{0}$ is the critical current of the Josephson junction, $I$ is the
bias current, $C$ is the junction capacitance, $\Phi_{0}=h/2e$ is the flux
quantum, $\hat{Q}$ denotes the charge operator and $\hat{\delta}$ represents
the gauge invariant phase difference across the junction, which obeys the
convectional quantum commutation relation $[\hat{\delta},\hat{Q}]=2ei$. The
states of the current-biased Josephson junction can be controlled through
the bias current $I(t)$ given by
\begin{equation}  \label{control}
I(t)=I_{dc}+\Delta I(t)=I_{dc}+ I_{\mu w}\cos\omega t,
\end{equation}
where the classical bias current is parameterized by a dc component $I_{dc}$
and an ac component with the magnitude $I_{\mu w}$ and frequency $\omega $.
For $I_{dc}<I_{0}$, the effective potential of the system (shown in
Fig.1(b)) has a series of metastable wells. At low temperature, the
current-biased junction has quantized energy levels, with the two lowest
energy states labeled as $|0\rangle $ and $|1\rangle$.
\begin{figure}
\centering
\includegraphics[width=3.3075in]{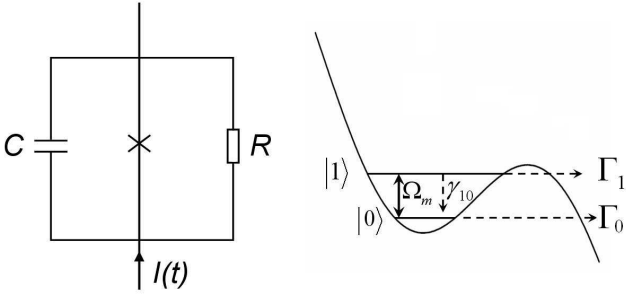}
\caption{(a) RCSJ equivalent circuit of a current-biased Josephson tunnel
junction. (b) Washboard potential of a current-biased Josephson junction
showing various coherent and incoherent processes at low temperature.
$|0\rangle $ and $|1\rangle $ are ground state and the first excited state
which are proposed to do quantum information process as a superconducting
phase qubit. }
\end{figure}
Microwaves induce transitions between $|0\rangle $ and $|1\rangle $ at a
frequency
\begin{equation}
\omega _{10}=\frac{E_{1}-E_{0}}{\hbar }=\omega _{p}(1-\frac{5}{36}\frac{%
\hbar \omega _{p}}{\Delta U}),  \label{energy seperation}
\end{equation}%
where $\omega _{p}(I_{dc})=2^{1/4}(2\pi I_{0}/\Phi
_{0}C)^{1/2}(1-I_{dc}/I_{0})^{1/4}$ is the small oscillation frequency at
the bottom of the washboard potential and $\Delta U(I_{dc})=(2\sqrt{2}%
I_{0}\Phi _{0}/3\pi )(1-I_{dc}/I_{0})^{3/2}$ is the barrier height. It is
apparent from Eq.(\ref{energy seperation}) that the energy spacing $\omega
_{10}$ is a function of the bias current $I_{dc}$. Therefore, if we ramp $%
I_{dc}$ from 0 to $I_{0}$, the barrier $\Delta U$ is decreasing. At certain $%
I_{dc}$ called switching current the system will tunnel out of the potential
well. In addition, a microwave with frequency matching the energy level
spacing will generate a transition between $|0\rangle $ and $|1\rangle $. As
shown in the top panel of Fig.2(b), the main peak of switching current
distribution corresponds to the tunneling from the ground state $|0\rangle $%
, and the resonant peak corresponds to the tunneling from the first excited
state $|1\rangle $. By plotting the frequency of microwave vs. the position
of the resonant peak we can obtain the energy spectrum of the junction.

To simulate the switching process of current-biased junction, we firstly
write the Hamiltonian of the junction in subspace \{$|0\rangle $, $|1\rangle
$\}
\begin{equation}
H_{JJ}=\hbar \left(
\begin{array}{cc}
0 & \Omega _{m}\cos \omega t \\
\Omega _{m}\cos \omega t & \omega _{10}(I_{dc})%
\end{array}%
\right) ,
\end{equation}%
where $\Omega _{m}=I_{\mu w}\sqrt{1/2\hbar \omega _{10}C}$ is Rabi
frequency. Considering the dissipative effect of environment, the time
evolution of the system can be described by the non-Hermitian effective
Hamiltonian
\begin{equation}
H_{eff}=H_{qb}-\frac{i\hbar }{2}(\gamma _{10}+\Gamma _{1})|1\rangle \langle
1|-\frac{i\hbar }{2}\Gamma _{0}|0\rangle \langle 0|,
\label{effective Hamiltonian}
\end{equation}%
where $\gamma _{10}$ is the energy relaxation rate from $|1\rangle $ to $%
|0\rangle $, and $\Gamma _{i}$ is the tunneling rate from state $|i\rangle $
(i = 0, 1) out of the potential (Fig. 1(b)). It is noticed that both $\gamma
_{10}$ and $\Gamma _{i}$ are functions of the bias current $I_{dc}$. At
temperature $T$, the relaxation rate $\gamma _{10}$ is given by\cite{Kopietz}
\begin{equation}
\gamma _{10}=\frac{\omega _{10}}{2\pi }\frac{R_{Q}}{R}[1+\coth (\frac{\hbar
\omega _{10}}{2k_{B}T})]\times |\langle 0|\hat{\delta}|1\rangle |^{2},
\end{equation}%
where $R_{Q}\equiv h/4e^{2}\simeq 6.45k\Omega $ is the natural quantum unit
of resistance and $R$ is the shunting resistance in RCSJ model (Fig. 1(a)).
The tunneling rate $\Gamma _{i}$ from the state $|i\rangle $ can be obtained
with the WKB method
\begin{equation}
\Gamma _{i}=\frac{1}{T(E_{i})}\exp (-\frac{2S_{f}(E_{i})}{\hbar }),
\label{Tunneling}
\end{equation}%
where $T(E_{i})$ is the classical period of motion and $S_{f}(E_{i})$ is the
action across the classically forbidden region.

\begin{figure}[tbp]
\centering
\includegraphics[width=3.5075in]{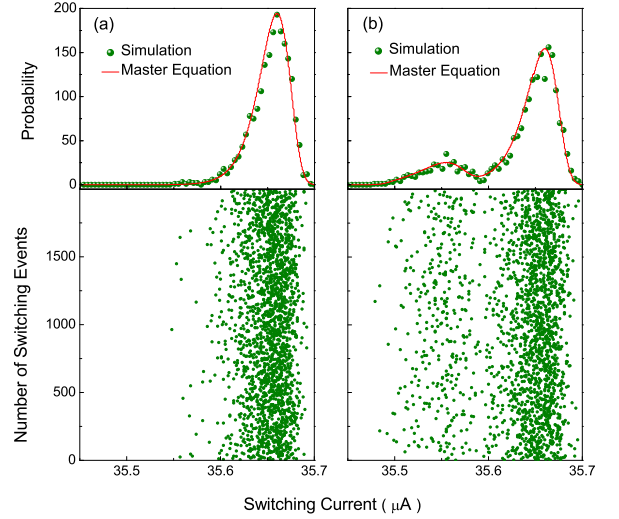}
\caption{(color online) Simulated switching currents (lower panel) of a
junction obtained by quantum jump approach (a) without microwave and, (b)
with microwave respectively. The parameters used for simulations are: $%
I_{0}=35.9\protect\mu $A, $C=4$pF, $\protect\omega /2\protect\pi $=9.02GHz,
and $\protect\gamma _{10}$=0.6$\protect\mu s^{-1}$. The ramping rate is $%
dI_{dc}/dt=4.5\times 10^{-3}\protect\mu $A/s. By making the histogram of the
switching currents we obtained the switching current distribution, shown as
the symbols in the top panel. The red lines are ensemble results obtained
using the master equations. }
\end{figure}

Then the quantum jump approach for simulating the switching process of
junctions can be summarized as follows:

(i) At $t=0$, initializing the junction in the ground state: $%
|\Psi(t=0)\rangle=|0\rangle$.

(ii) For $I_{dc}(t+\Delta t)=I_{dc}(t)+(dI/dt)\Delta t$, calculate the
corresponding energy spacing and various transition rates according to
Eq.(\ref{energy seperation}-\ref{Tunneling}).

(iii) Determine whether the system evolves according to the schr\"odinger
equation, or makes a 'jump'.\cite{Wen}

(a) If a quantum tunneling escape happens, register the switching current $%
I_s=I_{dc}(t)$, and then turn to step (v).

(b) If a relaxation event happens, then the system jumps to the ground state
$|0\rangle$.

(c) If no jumps happen, the system evolves under the influence of the
non-Hermitian form.\cite{Wen}

(iv) For the case (b) and (c), repeat from step (ii).

(v) Repeat to obtain the switching current $I_s$.

(vi) Average switching current $I_s$ over many simulation runs.

The numerical results obtained with quantum jump approach are shown in
Fig.2. The parameters we used in the simulation are from experiments.\cite{Yu2}
In addition, we calculate the switching current distribution with the
master equation method. The agreement between the quantum jump approach and
the master equation indicates that quantum jump approach is valid to model
the switching process of junction. Furthermore, as discussed in Sec. III,
quantum jump approach is more powerful than master equation method when
stochastic characteristics of a single quantum system play an important role.

\section{QUANTUM JUMP APPROACH TO SWITCHING PROCESS OF A JJ-TLS COUPLING
SYSTEM}

\begin{figure}
\centering
\includegraphics[width=3.3375in]{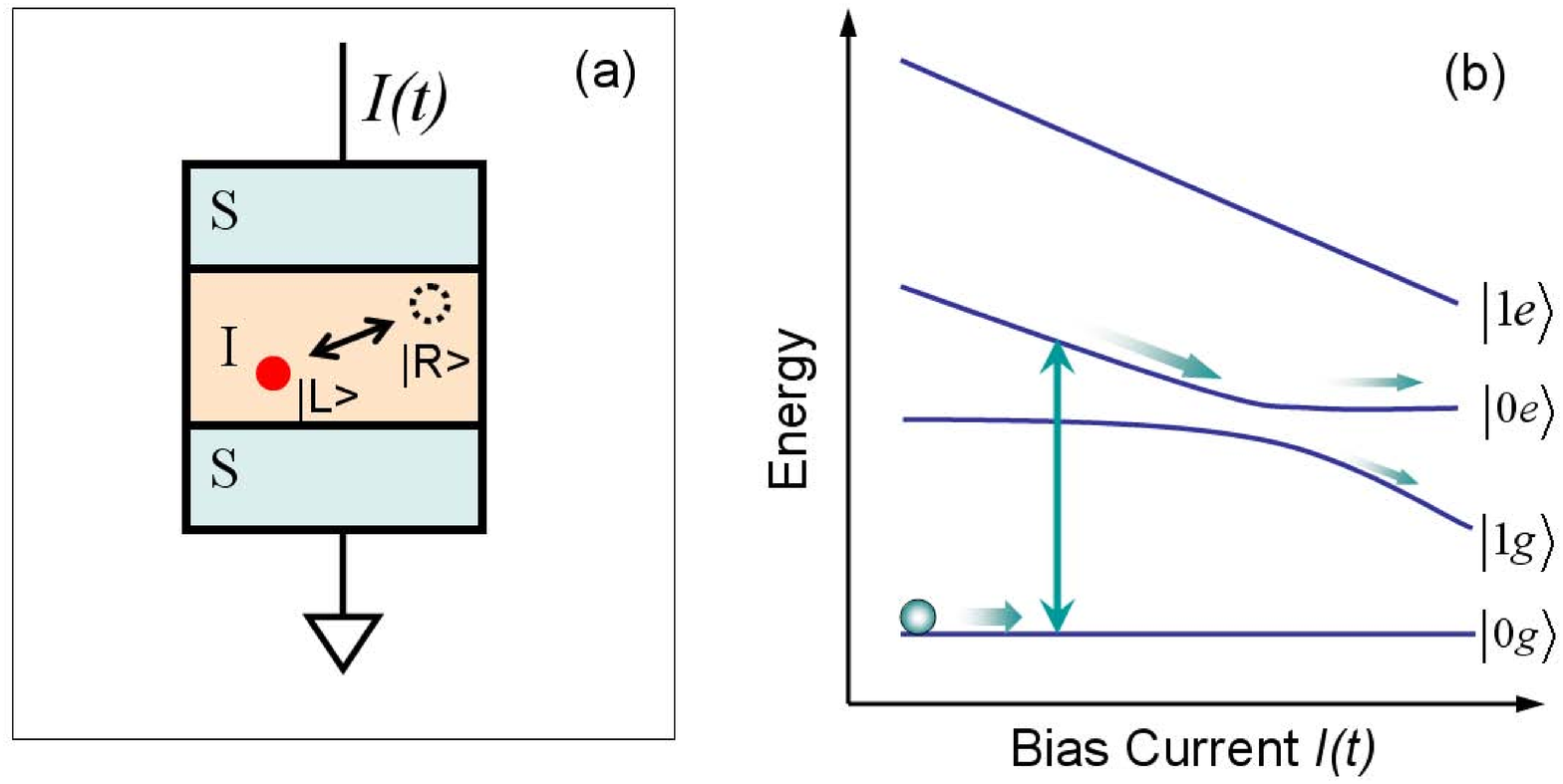}
\caption{(a) (Color online) Schematic of a TLS locating inside the Josephson
tunnel barrier. Some particles can tunnel between two lattice positions with
different wavefunctions $|L\rangle $ and $|R\rangle $, respectively. (b)
Illustration of transition from $|g\rangle $ to $|e\rangle $ for the TLS.
Assuming the system is initially in state $|0g\rangle $, at a certain biased
current $I_{dc}$, the microwave is resonant with $\protect\omega _{10}$ and
a transition from $|0g\rangle $ to $|1g\rangle $ happens. Furthermore, when
the biased current is ramped through the avoided energy level crossing, the
Landau-Zener transition may lead to a finite occupation probability in state $%
|0e\rangle $. }
\end{figure}

Firstly we give a brief description of the physics of JJ-TLS coupling
system. TLSs are extensively observed in superconducting phase\cite%
{Simmonds,Cooper,Martinis4}, charge\cite{Kim} and flux\cite{Lup} qubits
recently. A TLS is understood to be a particle or a small group of particles
that tunnels between two lattice configurations, with different wave
functions $|L\rangle $ and $|R\rangle $ corresponding to different junction
critical current $I_{0L}$ and $I_{0R}$, respectively (Fig. 3(a)). The
interaction Hamiltonian between the junction and TLS can be written as:\cite%
{Simmonds}
\begin{equation}
H_{int}=-\frac{\Phi _{0}I_{0R}}{2\pi }\cos \delta \otimes |R\rangle \langle
R|-\frac{\Phi _{0}I_{0L}}{2\pi }\cos \delta \otimes |L\rangle \langle L|.
\label{interaction Hamiltonian}
\end{equation}%
For convenience, we transfer to the energy eigenstate basis of TLS with $%
|g\rangle $ and $|e\rangle $ being the ground state and the excited state,
respectively. Then the total Hamiltonian of the JJ-TLS in the basis $%
\{|0g\rangle ,|1g\rangle ,|0e\rangle ,|1e\rangle \}$ is given by:\cite%
{Wen,Clare}
\begin{equation}
H=\hbar \left(
\begin{array}{cccc}
0 & \Omega _{m}\cos \omega t & 0 & 0 \\
\Omega _{m}\cos \omega t & \omega _{10}(I_{dc}) & \Omega _{c} & 0 \\
0 & \Omega _{c} & \omega _{TLS} & \Omega _{m}\cos \omega t \\
0 & 0 & \Omega _{m}\cos \omega t & \omega _{10}(I_{dc})+\omega _{TLS}%
\end{array}%
\right) ,  \label{origin Hamiltonian}
\end{equation}%
where $\omega _{TLS}$ is the energy frequency of the TLS, and $\Omega _{c}$
is the coupling strength between Josephson junction and TLS. In experiments, the
coupling strength $\Omega _{c}$ can be characterized in spectroscopic
measurements and usually lies from 20MHz to 200MHz.\cite%
{Yu2,Simmonds,Cooper,Martinis4,Kim,Lup} The time evolution of the JJ-TLS
coupling system under the dissipative effect of environments can be
described by the effective Hamiltonian
\begin{eqnarray}
H_{eff} &=&H-\frac{i\hbar }{2}\Gamma _{0g}|0g\rangle \langle 0g|-\frac{%
i\hbar }{2}(\gamma _{10}+\Gamma _{1g})|1g\rangle \langle 1g|  \nonumber \\
&&-\frac{i\hbar }{2}\Gamma _{0e}|0e\rangle \langle 0e|-\frac{i\hbar }{2}%
(\gamma _{10}+\Gamma _{1e})|1e\rangle \langle 1e|,
\end{eqnarray}%
where $\Gamma _{i}$ is the tunneling rate from state $|i\rangle $. We
emphasize that in the asymmetric double well model of TLS, the energy basis
of TLS is approximated to the position basis. In this approximation, states $%
|g\rangle $ and $|e\rangle $ correspond to different critical currents.
Therefore, the tunneling rates from different states are different. With no
loss of generality, suppose the state $|e\rangle $ corresponds to the
smaller critical current. Then the procedure for simulating the switching
process of the JJ-TLS coupling system can be summarized as follows:

(i) Initializing the system in state $|0g\rangle$ for $flag=0$, or in state $%
|0e\rangle$ for $flag=1$, where $flag$ is a marker.

(ii) For $I_{dc}(t+\Delta t)=I_{dc}(t)+(dI/dt)\Delta t$, calculate the
corresponding energy spacing $\omega_{10}(I_{dc})$ and various transition
rates.

(iii) Determine whether the system evolves according to the schr\"odinger
equation, or makes a 'jump'.

(a) If a quantum tunneling event happens, register the switching current $%
I_s=I_{dc}(t)$. Furthermore, if the system tunnels from $|0g\rangle$ or $%
|1g\rangle$, set $flag=0$; else if the system tunnels from $|0e\rangle$ or $%
|1e\rangle$, set $flag=1$; and then turn to step (v).

(b) If a relaxation event happens, then the system jumps to the
corresponding ground state $|0g\rangle$ or $|0e\rangle$, i.e., the system
jumps from $|1g\rangle$ to $|0g\rangle$, or from $|1e\rangle$ to $|0e\rangle$
.

(c) If no jumps happen, the system evolves under the influence of the
non-Hermitian form.

(iv) For the case (b) and (c), repeat from step (ii).

(v) Repeat to obtain the switching current $I_s$.

\begin{figure}
\centering
\includegraphics[width=3.3075in]{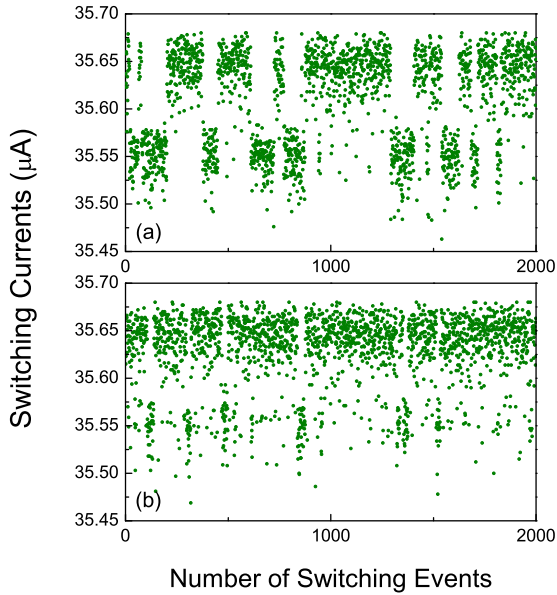}
\caption{Simulated trajectories of switching current of JJ-TLS coupling
system under different microwave amplitude. The parameters used in the
simulations are: $\protect\omega /2\protect\pi $=9.02GHz, $\protect\omega %
_{TLS}/2\protect\pi =8.7$GHz, $\Omega _{c}/2\protect\pi =200$MHz, $\protect%
\gamma _{10}$=0.6$\protect\mu s^{-1}$, $dI_{dc}/dt=4.5\times 10^{3}\protect%
\mu $A/s and (a) $\Omega _{m}$=2MHz, (b) $\Omega _{m}$=10MHz. With the
microwave amplitude increasing , the jumps between upper branch and lower
branch become more frequent, resulting in a shorter lifetime in each branch.
}
\end{figure}
The simulation results are shown in Fig. 4. It is apparent that the
switching current jumps between upper branch and lower branch randomly,
which is the major characteristic of macroscopic quantum jumps observed in
experiments.\cite{Yu2} In addition, it is found that the jumps become more
frequent with the microwave amplitude increasing (Fig.4). The underlying
physics can be understood as follows. The jumps between upper branch and
lower branch of the switching current are fulfilled through the coupling
between state $|1g\rangle $ and $|0e\rangle $. With increasing the microwave
amplitude, the system initialized in $|0g\rangle $ has a larger transition
rate to $|1g\rangle $ in the expression
\begin{equation}
\Gamma =\frac{\Omega _{m}^{2}\gamma }{2(\Delta ^{2}+\gamma ^{2})},
\end{equation}%
where $\gamma =(\gamma _{10}+\Gamma _{0g}+\Gamma _{1g})/2$ and $\Delta
=\omega _{10}-\omega $. Therefore, it is much easier for the system to jump
to state $|0e\rangle $, i.e., jump from the upper branch to the lower
branch, and vice versa. It is easier to understand for the extreme case $%
\Omega _{m}=0$. Then the system has no probability to occupy state $%
|1g\rangle $, thus no probability to transfer to $|0e\rangle $.

Furthermore, it is noticed that the transition process from $|1g\rangle $ to
$|0e\rangle $ is actually a Landau-Zener transition as illustrated in Fig.
3(b). Disregarding all decay terms, the asymptotic probability of a
Landau-Zener transition is given by\cite{Landau,Zener}
\begin{equation}
P_{LZ}=\exp (-2\pi \frac{\hbar \Omega _{c}^{2}}{\upsilon }),  \label{Landau}
\end{equation}
where $2\hbar \Omega _{c}$ is the magnitude of the energy splitting, and $%
\upsilon \equiv d\varepsilon /dt$ denotes the variation rate of the energy
spacing for noninteracting levels. Notice that $\upsilon \equiv
(d\varepsilon /dI_{dc})(dI_{dc}/dt)$, where $d\varepsilon /dI_{dc}$ is
determined by the intrinsic parameters of the junction and $dI_{dc}/dt$ is
determined by the ramping rate. It can be easily inferred from Eq.(\ref{Landau})
that as the ramping rate increased, the transition rate between
$|1g\rangle $ and $|0e\rangle $ becomes smaller. Therefore, the jumps between
upper branch and lower branch of the switching currents become less frequent
and the lifetime for each branch is longer. To support this argument, we
simulate the trajectories of the switching currents for different ramping
rates. As shown in Fig. 5, with the ramping rate increasing , the jumps
between different branches become less frequent, as expected from our
theoretical analysis.

\begin{figure}
\centering
\includegraphics[width=3.3075in]{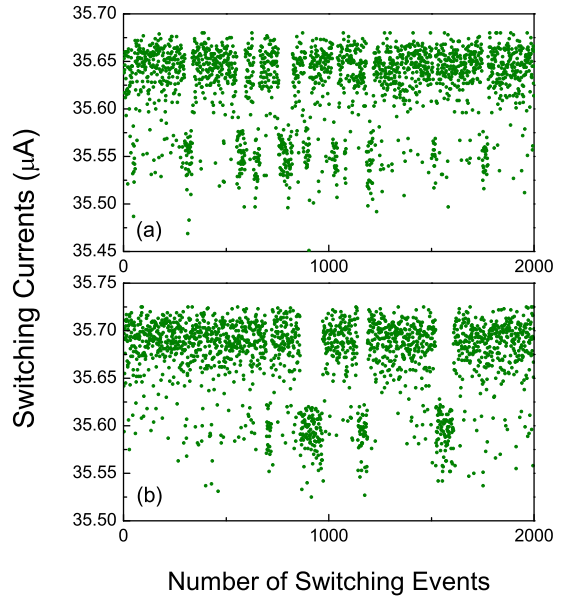}
\caption{Simulated trajectories of switching current of JJ-TLS coupling
system under different bias current ramping rates. The parameters used in
the simulations are: $\protect\omega /2\protect\pi $=9.02GHz, $\protect%
\gamma _{10}$=0.6$\protect\mu s^{-1}$, $\Omega _{m}$=10MHz and (a) $%
dI_{dc}/dt=4.5\times 10^{3}\protect\mu $A/s, (b) $dI_{dc}/dt=8.0\times 10^{3}%
\protect\mu $A/s. With the ramping rate increasing , the jumps between upper
branch and lower branch become less frequent, resulting in a longer lifetime
in each branch. In addition, the switching current in both branches becomes
higher for the larger ramping rate. }
\end{figure}

\section{EXPERIMENTS}

\begin{figure}
\centering
\includegraphics[width=3.3075in]{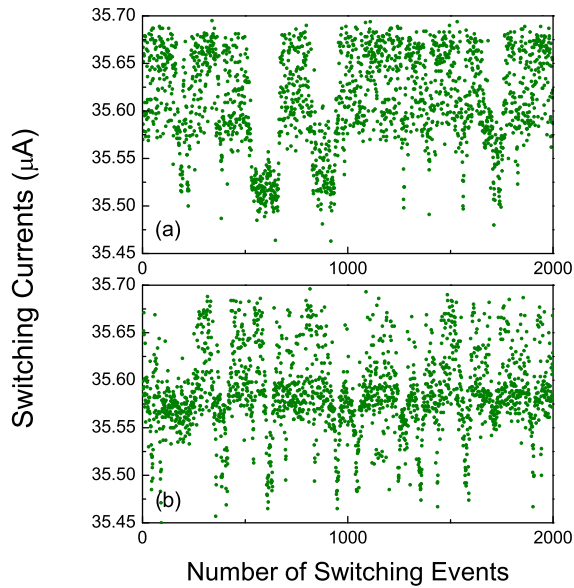}
\caption{Experimental trajectories of switching current of JJ-TLS coupling
system under different microwave power with (a) -7.2dBm and (b) -5.8dBm. The
ramping rate is $dI_{dc}/dt=4.5\times10^{3}\protect\mu$A/s. As increasing
the microwave power, the jumps become much more frequent. }
\end{figure}

We have compared the results of our theoretical analysis with the
experimental data. The sample used in our experiments was a 10$\mu $m$\times
$ 10$\mu $m Nb/Al$O_{x}$/Nb Josephson junction. The junction parameters are $%
I_{0}\approx 36\mu $A and $C\approx 4$pF, respectively. The device was
thermally anchored to the mixing chamber of a dilution refrigerator with a
base temperature of about 18mK. Additional three-layer mu-metal surrounding
the dewar was used to shield the magnetic field. All electrical leads that
connect the junction to room temperature electronics were carefully filtered
by resistor-capacitor (RC) filters and copper powder filters. The center
conductor of an open-ended coaxial cable was placed above the junction for
application of microwave. This arrangement resulted in $>$110dB attenuation
between the end of the coaxial cable and the junction. A saw-tooth bias
current was applied with a repetition rate of 30Hz $\sim $ 300Hz.\cite{Yu2,Dong}
The junction voltage was amplified by a differential amplifier
and the switching current was recorded when a voltage greater than the
threshold was first detected during every ramp.
\begin{figure}
\centering
\includegraphics[width=3.3075in]{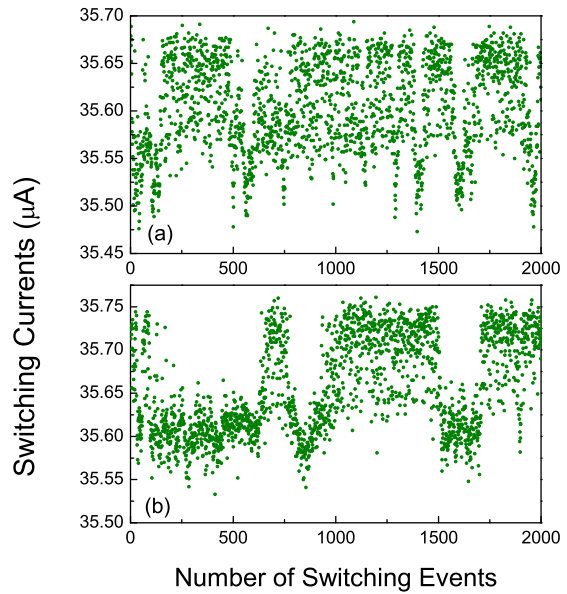}
\caption{Experimental trajectories of switching current of JJ-TLS coupling
system under different ramping frequency. The microwave power is -6.6dBm and
the ramping rates are (a) $dI_{dc}/dt=4.5\times 10^{3}\protect\mu $A/s and
(b) $dI_{dc}/dt=8.0\times 10^{3}\protect\mu $A/s, respectively. With the
ramping rate increasing, the jumps between different branches become less
frequent. }
\end{figure}

In the spectroscopy measurement of the junction, an avoided crossing caused
by the coupling between junction and TLS was observed at $\omega /2\pi =8.7$
GHz with an energy splitting $2\Omega _{c}/2\pi =400$MHz.\cite{Yu2} When a
microwave field with $\omega /2\pi =9.02$ GHz was applied, the coupling
between junction and TLS was turned on. In this case the macroscopic quantum
jumps between upper branch and lower branch of the switching current were
observed. To investigate the effect of microwave power as
discussed in Sec. III, we fixed the microwave frequency at $\omega =9.02$
GHz and the ramping rate at $dI_{dc}/dt=4.5\times 10^{3}$ $\mu $A/s. The
microwave power was adjusted from -20dBm to -3dBm. As shown in Fig. 6, with
the microwave power increasing, quantum jumps between different branches
become much frequent. Similarly, to investigate the effect of ramping rate,
we adjusted the ramping rate $dI_{dc}/dt$ from $2.0\times 10^{3}$ $\mu $A/s
to $16.0\times 10^{3}$ $\mu $A/s while keeping other parameters fixed. As
expected, the increasing of ramping rate results in less frequent jumps
between upper branch and lower branch of the switching current (Fig. 7). The
agreement between our simulated results and the experimental data confirmed
the validity of the quantum jump approach.

\section{CONCLUSION}

We have used the quantum jump approach to simulate the switching process of
the JJ-TLS coupling system. The mechanism that dominates the quantum jumps
phenomenon was discussed. In addition, we investigated the parameters that
have effects on the behavior of quantum jumps. It is found that a higher
microwave power or a smaller ramping rate can make the quantum jumps happen
more frequently, which has significance in controlling the state of TLS.
Furthermore, our theoretical results agree with the experimental data,
indicating the validity of our approach. The model and method we used here
can be easily generalized to other solid-state systems such as flux and
charge qubits, quantum dots, trapped irons and so on.

\section{ ACKNOWLEDGMENTS}

This work was partially supported by the NSFC (under Contracts No. 10674062
and No. 10725415), the State Key Program for Basic Research of China (under
Contract No. 2006CB921801), and the Doctoral Funds of the Ministry of
Education of the People's Republic of China (under Contract No.
20060284022).

\end{document}